\documentclass[aip,jcp,reprint,superscriptaddress,floatfix]{revtex4-1}
\usepackage[utf8]{inputenc}
\usepackage{color}
\usepackage{graphicx}

\setcitestyle{open={},close={}}

\newcommand{\cotc}{{[Co(TC-3,3)(NO)]}}

\begin{document}
	\title{Multi-reference perturbation theory with Cholesky decomposition for the density matrix renormalization group}
	\author{Leon Freitag}
	\email{leon.freitag@phys.chem.ethz.ch}
	\affiliation
	{ETH Zürich, Laboratorium für Physikalische Chemie,
	Vladimir-Prelog-Weg 2, 8093 Zürich, Switzerland
	}
	\author{Stefan Knecht}
	\email{stefan.knecht@phys.chem.ethz.ch}
	\affiliation
	{ETH Zürich, Laboratorium für Physikalische Chemie,
	Vladimir-Prelog-Weg 2, 8093 Zürich, Switzerland
	}
	\author{Celestino Angeli}
	\email{anc@unife.it}
	\affiliation
	{Dipartimento di Scienze Chimiche e Farmaceutiche,
	Universit\`{a} di Ferrara, Via Fossato di Mortara 17,
	44121 Ferrara, Italy
	}
	\author{Markus Reiher}
	\email{markus.reiher@phys.chem.ethz.ch}
	\affiliation
	{ETH Zürich, Laboratorium für Physikalische Chemie,
	Vladimir-Prelog-Weg 2, 8093 Zürich, Switzerland
	}

	\begin{abstract}
		We present a second-order \textit{N}-electron valence state perturbation theory (NEVPT2) 
based on a density matrix renormalization group (DMRG) reference wave function that exploits a 
Cholesky decomposition of the two-electron repulsion integrals (CD-DMRG-NEVPT2). 
With a parameter-free multireference perturbation theory approach at hand, the latter allows us 
to efficiently describe static and dynamic correlation in large molecular systems. 
We demonstrate the applicability of CD-DMRG-NEVPT2 for spin-state energetics of 
spin-crossover complexes involving calculations with more than 1000 atomic basis functions. 
We first assess in a study of a heme model the accuracy of the strongly- and partially-contracted 
variant of CD-DMRG-NEVPT2 before embarking on resolving a controversy about the spin ground state 
of a cobalt tropocoronand complex.
	\end{abstract}
	\maketitle
	
	\section{Introduction}
	An electronic structure that is governed by strong electron correlation effects is a commonly encountered 
phenomenon in molecules that are, for example, (i) in a non-equilibrium structure, (ii) in an electronically excited
state and/or (iii) contain transition metal center(s). 
Multiconfigurational methods, which have been very successful in describing strongly-correlated systems\cite{szal12,Roca-Sanjuan_WIREsComputMolSci_Multiconfiguration_2012}, 
typically feature a separation of the electron correlation into a static and a dynamic contribution 
\cite{Bartlett__Applications_1994,Sinanoglu_JChemPhys_ManyElectron_1963,Boguslawski_JPhysChemLett_Entanglement_2012}. 
Static correlation is often described by a complete active space self-consistent field 
(CASSCF) \textit{ansatz}\cite{Roos_ChemPhys_complete_1980,olse11}, which requires a careful 
selection of a limited number of (partially occupied) \emph{active orbitals}
that may be automatized \cite{Stein_JChemTheoryComput_Automated_2016,stein16b}.
As the computational cost for CASSCF scales exponentially with the number of active orbitals, 
CASSCF calculations (on traditional computer architectures) are limited to about 18 electrons in 18 orbitals.\cite{Aquilante_JComputChem_Molcas_2015}
By contrast, the density matrix renormalization 
group\cite{White_PhysRevLett_Density_1992,White_PhysRevB_Densitymatrix_1993,Schollwock_AnnPhys_densitymatrix_2011} (DMRG)
approach in quantum chemistry\cite{Legeza2008,Chan_Introduction_2008,Marti_ZPhysChem_Density_2010,Marti2011,Chan2011a,
wout14,Kurashige2014,Yanai2015,szalay2015tensor,Chan_JChemPhys_Matrix_2016} 
in combination with a self-consistent-field orbital optimization \textit{ansatz} (DMRG-SCF) 
is capable of approximating CASSCF wave functions to arbitrary accuracy with a polynomial 
rather than an exponential scaling. DMRG-SCF therefore provides access to much larger active orbital spaces 
than those that are in reach for standard CASSCF.

For a quantitative description of electron correlation a subsequent step must account 
also for dynamic correlation. In this context, multireference perturbation theories such as the second-order 
complete active space perturbation theory (CASPT2)\cite{Andersson_JPhysChem_CASPT2_1990,Andersson_JChemPhys_Secondorder_1992} 
or \textit{N}-electron valence state perturbation theory (NEVPT2)\cite{Angeli_JChemPhys_Introduction_2001} have 
been successfully employed in obtaining energies, properties and approximate wave functions for a variety of 
strongly-correlated systems.
\cite{Pierloot_MolPhys_CASPT2_2003,Havenith_JChemPhys_Calibration_2004,Pulay_IntJQuantumChem_perspective_2011,Roca-Sanjuan_WIREsComputMolSci_Multiconfiguration_2012}

Both CASPT2\cite{Andersson_JPhysChem_CASPT2_1990} and NEVPT2\cite{Angeli_JChemPhys_nelectron_2002} require 
the evaluation of higher-order reduced density matrices (RDMs) in the active space. The salient properties of
NEVPT2 (see below) requires the four-body RDM, which scales as $L^8$ 
where $L$ is the number of active orbitals. This $L^8$ scaling, although sub-exponential, still puts a constraint 
on the active orbital space size tractable by multireference perturbation theory, in particular in combination 
with a DMRG reference wave function. Therefore, in addition to straightforward implementations such 
as DMRG-CASPT2\cite{Kurashige_JChemPhys_Secondorder_2011} or strongly-contracted 
(SC-) DMRG-NEVPT2\cite{Guo_JChemTheoryComput_NElectron_2016,knec16}, several approaches have been proposed 
to tackle the scaling problem of the higher-order RDMs. The CASPT2 implementation of \citet{Kurashige_JChemPhys_Complete_2014} and the NEVPT2 implementation 
of \citet{Zgid_JChemPhys_study_2009} employ cumulant-type expansions to approximate the three- and four-body RDMs which, however,
may entail $N$-representability problems of the approximated higher-order RDMs and/or lead to numerical instabilities in
the perturbation summations. Very promising alternative formulations such as matrix product state perturbation 
theory (MPS-PT)\cite{Sharma_JChemPhys_Communication_2014,shar16}, a time-dependent formulation\cite{Sokolov_JChemPhys_timedependent_2016} 
or the projected approximation to SC-NEVPT2\cite{Roemelt_JChemPhys_projected_2016} avoid the construction of higher-order RDMs altogether.

Besides multireference perturbation theory, DMRG-SCF has been
successfully combined with other methods such as multireference
configuration interaction
(MRCI)\cite{Saitow_JChemPhys_Multireference_2013}, canonical
transformation theory,\cite{Yanai_JChemPhys_Multireference_2010,Neuscamman_JChemPhys_Strongly_2010,Yanai_PhysChemChemPhys_Extended_2012}
coupled cluster (CC) theory, \cite{Veis_Coupled_2016} 
and short-range density functional theory \cite{hede15}
to describe dynamic correlation.

In this work, we present a full-fledged NEVPT2 for a DMRG reference function that exploits density fitting for the two-electron
integrals in order to make large molecules more accessible to this methodology.
We apply our implementation to a prototypical problem in transition metal chemistry, namely the spin state energetics problem\cite{cost13}.
In computational transition metal and bioinorganic chemistry, gaining insight into the electronic structure, spin state energetics and 
reactivity of (polynuclear) transition metal complexes with extensive ligands are key components for the
understanding of the chemistry of metalloproteins\cite{swart15}. 
Albeit being in many cases unreliable and/or heavily dependent on the choice of the density functional, 
density functional theory (DFT) is the standard approach to study such systems\cite{Cramer_PhysChemChemPhys_Density_2009,Tsipis_CoordChemRev_DFT_2014}, 
mainly because of its affordable computational cost. 
Thermal spin crossover complexes (SCOs) constitute a prime example for the difficulty of DFT with present-day functionals. On the one hand, this may be due 
to static correlation abundant in many transition metal systems\cite{Harvey_AnnuRepProgChemSectC_accuracy_2006} which 
none of the currently available standard density functionals can describe properly. On the other hand, the prediction of the correct 
ground state in SCOs is often a matter of a few kJ/mol, an accuracy which DFT does not always offer.\cite{Swart_Accurate_2010} 
To this end, multiconfigurational studies on SCOs have become increasingly popular.\cite{Sousa_Initio_2015} 
	 
In the quest for making multiconfigurational methods capable of treating larger molecular systems of arbitrary complexity 
(and thus establishing them as a part of the standard toolbox in theoretical transition metal and bioinorganic chemistry), providing access to larger
active orbital spaces only partly solves the problem as the convergence of the results can be slow with 
respect to the size of the atomic basis set. For large atomic basis sets, the 
transformation of the two-electron repulsion integrals from an atomic orbital (AO) 
to a molecular orbital (MO) basis required for the subsequent perturbation theory step will become a (second) bottleneck. 
In this context, a promising \emph{ansatz} has been recently proposed by Neese and co-workers\cite{Guo_JChemPhys_SparseMapsA_2016}, 
who have developed a domain-based pair local natural orbital NEVPT2 (DLPNO-NEVPT2) formulation.
Very recently, Evangelista and co-workers \cite{lich15,hann16} put forward a novel second-order multireference perturbation theory based on the driven similarity renormalization group which exploits factorized two-electron integrals to enable an on-the-fly generation of the two-electron repulsion integrals in the MO basis while avoiding any explicit storage of the latter.

In this work we will take advantage of a Cholesky decomposition (CD) of the two-electron 
integral matrix\cite{Beebe_IntJQuantumChem_Simplifications_1977,Koch_JChemPhys_Reduced_2003}. Although the idea of CD for two-electron integrals dates 
back to 1977\cite{Beebe_IntJQuantumChem_Simplifications_1977}, the approach has been only recently 
fully explored and elaborated by Aquilante, Pedersen, Lindh and co-workers\cite{Aquilante_Cholesky_2011}. They also 
implemented CD for traditional multiconfigurational electronic structure methods such as 
CASSCF\cite{Aquilante_JChemPhys_Accurate_2008,Delcey_JChemPhys_Analytical_2014} and 
CASPT2 \cite{Aquilante_JChemTheoryComput_Cholesky_2008,Aquilante_JChemPhys_Systematic_2009,
Roca-Sanjuan_WIREsComputMolSci_Multiconfiguration_2012}. 
While these developments have spurred significantly various applications of CASPT2 in theoretical inorganic 
and bioinorganic chemistry\cite{Roca-Sanjuan_WIREsComputMolSci_Multiconfiguration_2012}, 
CASPT2 in its most successful formulation contains a parameter (the so-called IPEA shift) with a default 
value of 0.25 a.u.~introduced by \citet{Ghigo_ChemPhysLett_modified_2004} to match a set 
of experimental dissociation energies of main group diatomics. The default value of the IPEA shift 
has sparked a controversy among researchers employing CASPT2 for problems other than dissociation 
energies, such as magnetic coupling constants and spin-state energetics of SCOs\cite{Queralt_JComputChem_applicability_2008,Kepenekian_JChemPhys_What_2009,
LawsonDaku_JChemTheoryComput_Accurate_2012,Vela_JComputChem_zerothorder_2016} which still remains unsettled. 
Moreover, CASPT2 shows in some cases unstable results due to the so-called 'intruder-state problem' due to the appearance of very small
denominators in the perturbation expansion. This problem has been counteracted
with a level-shift technique\cite{roos95,fors97}, 
but the final CASPT2 energies depend on the value of the level shift.
By contrast, NEVPT2, which is parameter-free and has the
noteworthy property to avoid the intruder-state problem (owing to the
sophistication of its zero order Hamiltonian), not only avoids any IPEA-shift controversy but has also shown promising
results for spin-state energetics in model compounds and SCOs\cite{Havenith_JChemPhys_Calibration_2004,Queralt_JComputChem_applicability_2008,Vela_JComputChem_zerothorder_2016}

We here present a DMRG-NEVPT2 implementation employing CD that is both capable 
of treating large active orbital spaces \text{and} large AO basis sets which allows for 
multireference calculations on large transition metal complexes that are typically encountered in bioinorganic chemistry. 
To demonstrate its capabilities we examine the spin-state energetics of two prime examples of
SCOs.

\section{Theory and Computational Methodology}

As details of the NEVPT2 approach, quantum-chemical DMRG and its combination to DMRG-NEVPT2 have been discussed in
detail
elsewhere\cite{Angeli_JChemPhys_Introduction_2001,Angeli_ChemPhysLett_Nelectron_2001,Angeli_JChemPhys_nelectron_2002,Zgid_JChemPhys_study_2009,White_PhysRevLett_Density_1992,White_PhysRevB_Densitymatrix_1993,White_JChemPhys_initio_1999,Chan_Introduction_2008,Marti_ZPhysChem_Density_2010,Schollwock_AnnPhys_densitymatrix_2011,wout14,szalay2015tensor,Guo_JChemTheoryComput_NElectron_2016,Chan_JChemPhys_Matrix_2016},
we only outline briefly the Cholesky decomposition and show how it is employed in the context of CD-DMRG-NEVPT2.
	
	\subsection{Cholesky Decomposition}
	Mathematically, CD\cite{Benoit_BullGeodesique_Note_1924} is a special case of a LU decomposition of a positive semidefinite symmetric matrix $\mathbf{M}$ into a product of a triangular matrix $\mathbf{L}$ and its transpose,
	\begin{equation}
		\mathbf{M}=\mathbf{L L}^T .
	\end{equation}
	$\mathbf{M}$ now contains two-electron integrals $(ij|kl)$ (with indices $ij$ and $kl$ combined to row and column indices of $\mathbf{M}$). 
Of particular interest is the incomplete CD, where $\mathbf{M}$ may be approximated to an arbitrary accuracy with \emph{Cholesky vectors} $\vec{L}^J$, which constitute the columns of the matrix $\mathbf{L}$\cite{Aquilante_Cholesky_2011},
	\begin{equation}
		\mathbf{M}\approx \sum_{J=1}^M \vec{L}^J \otimes (\vec{L}^J)^T ,
	\end{equation}
	where $M$ is typically significantly smaller than the full dimension of $\mathbf{M}$. One important advantage of CD is that the Cholesky vectors can be computed without evaluating the full matrix $\mathbf{M}$\cite{Aquilante_Cholesky_2011}, resulting in large disk space savings. Two-electron integrals may then be reconstructed from the corresponding Cholesky vector elements,
	\begin{equation}
		(ij|kl) \approx \sum_{J=1}^M L_{ij}^J L_{kl}^J  , \label{eqn:cholesky}
	\end{equation}

	CD also proves particularly useful in integral transformations: instead of a full-fledged integral transformation, we
only need to transform Cholesky vectors,
	
	\begin{equation}
		L_{\mu\nu}^J = \sum_i\sum_j c_{\mu i}c_{\nu j} L_{ij}^J ,
	\end{equation}
	where $\mu,\nu$ and $i,j$ refer to indices in the AO and MO basis, respectively. 
The cost of such a transformation is $K^3M$ in contrast to a formal $K^5$ scaling for the full integral transformation. 
In NEVPT2, two-electron integrals with up to two inactive and two virtual orbital indices must be transformed from AO to the MO basis, 
which is typically the most time consuming step following the calculation of four-body RDMs. 
Additionally, in a multi-state calculation integrals with inactive/virtual indices must be transformed to a state-specific representation. 
In our implementation of CD-DMRG-NEVPT2 the integral transformation based on Cholesky vectors yielded 
by far the largest computational savings. Additionally, we implemented the construction of the 
Fock matrix directly from the Cholesky vectors (albeit without employing the local 
K screening by \citet{Aquilante_JChemPhys_Lowcost_2007} for the exchange contributions),
 however the computational advantage arising from this step is negligible compared to the integral transformation.

\subsection{Selected Spin-Crossover Complexes}
	
\begin{figure}[h]
		\centering
		\includegraphics[width=0.35\textwidth]{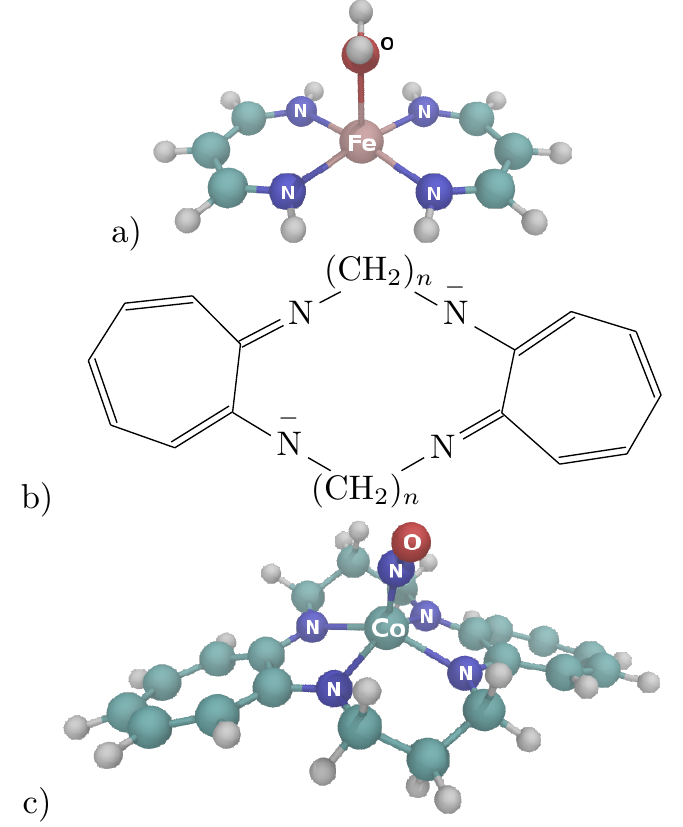}
		\caption{Structures of the compounds used in this work: a) {[Fe(C$_3$N$_2$H$_5$)$_2$(OH$_2$)]} (\emph{model 2}), b) the TC($n$,$n$) ligand, c) \cotc{}}\label{fig:structures}
\end{figure}

The first complex studied in this work, {[Fe(C$_3$N$_2$H$_5$)$_2$(OH$_2$)]} (Fig.~\ref{fig:structures}a) 
is a heme model employed by \citet{Strickland_JPhysChemB_SpinForbidden_2007} (named \emph{model 2} in 
their paper). We employ the \emph{model 2} complex as a benchmark system and
 assess the performance of the CD-DMRG-NEVPT2 method for spin-state energetics by comparing the relative 
energies of the lowest singlet, triplet and quintet states of the \emph{model 2} with previous 
accurate CCSD(T) calculations. 	

The second compound investigated is a cobalt tropocoronand nitrosyl complex, \cotc{} (Fig.~\ref{fig:structures}c). 
In a series of homologous tropocoronand complexes with an increasing methylene chain length $n$, initially \cotc{} with $n$=3 was earlier 
found to be paramagnetic unlike its homologs with larger $n$\cite{Ellison_InorgChem_TiltAsymmetry_1998,
Franz_InorgChem_Pentacoordinate_2001,Jaworska_ChemPhys_Theoretical_2007}. However, a recent DFT and 
experimental study\cite{Hopmann_InorgChem_SingletTriplet_2015} showed the opposite. 
Nitrous oxide (NO) is a well-known non-innocent ligand\cite{Jorgensen_CoordChemRev_Differences_1966,
Kaim_EurJInorgChem_Shrinking_2012}, and therefore transition metal nitrosyl complexes are known to have an intricate and 
challenging electronic structure, described best with multiconfigurational methods.
\cite{Radon_JPhysChemA_Binding_2008,Radon_JPhysChemB_Electronic_2010,bogu11,Boguslawski_JChemTheoryComput_Accurate_2012,
Freitag_PhysChemChemPhys_Orbital_2015} Here, we employ CD-DMRG-NEVPT2 to contribute 
to the discussion of spin state energetics of \cotc{}.

	\subsection{Computational Details}
	The structures of both complexes studied in this work were taken from previous publications: the \emph{model 2} structures were optimized by \citet{Strickland_JPhysChemB_SpinForbidden_2007} for each individual spin state with the B3LYP\cite{Becke_JChemPhys_B3_1993,Lee_PhysRevB_Development_1988} density functional, a Los Alamos effective core potential and LACV3P basis set for Fe and the 6-311G* basis set for other atoms with a $C_{2v}$ symmetry constraint. The \cotc{} structures were taken from Ref.~\citenum{Hopmann_InorgChem_SingletTriplet_2015}, where they were optimized with the PW91\cite{Perdew_PhysRevB_Atoms_1992} density functional and 6-311G** basis set. 
	
	Subsequently, DMRG-SCF calculations employing a 14 electrons in 18 spatial orbitals, i.e., (14,18), active space for the \emph{model 2} complex and a
	(22,22) active space for \cotc{} have been performed. A procedure following our automated active space selection\cite{Stein_JChemTheoryComput_Automated_2016} delivered the choice of the active spaces (more details on the active spaces and the selection procedure can be found in the Supplementary Information). 
To study the effect of the number of renormalized block states $m$ on the results and the errors arising from the CD approximation, different approaches to prepare the DMRG wave function were investigated for the \emph{model 2} complex: i) DMRG-SCF calculations with $m$ values of 256 and 512; ii) DMRG-SCF calculations with $m=256$ followed by a DMRG-CI calculation with $m=512$ and $m=1024$ (denoted as 512/256 and 1024/256, respectively, in the following), and iii) DMRG-SCF calculations with $m$ values of 256 but without employing the Cholesky decomposition, denoted as $256$* in the following. For the \cotc{} complex, only DMRG-SCF calculations with $m=512$ were carried out.
	
We have implemented the CD-DMRG-NEVPT2 approach in a modularized version of the original QDNEVPT2 program by Angeli and co-workers
\cite{Angeli_JChemPhys_quasidegenerate_2004}\ for both the strongly contracted (SC) and
partially contracted (PC) variant of NEVPT2\cite{Angeli_JChemPhys_nelectron_2002}. 
	SC- and PC-NEVPT2 calculations were performed with all DMRG-SCF and DMRG-CI reference wave functions. The ANO-RCC\cite{Roos_JPhysChemA_New_2005} basis set was chosen for all calculations: in the \emph{model 2} complex, a moderately sized  double-zeta (VDZP) contraction on all atoms was employed to facilitate the comparison with a conventional integral calculation (yielding 248 basis functions in total), whereas for \cotc{} a larger triple-zeta (VTZP) contraction on all atoms, totaling 1147 basis functions, was used. 
	
	Additionally, CCSD(T) calculations were performed on the \emph{model 2} complex employing 
the same ANO-RCC-VDZP basis set for comparison. For the open-shell species, restricted
 open-shell reference wave functions and the triples contributions 
according to \citet{Watts_JChemPhys_Coupledcluster_1993} were calculated.
	
	
Cholesky vectors in the AO basis were generated based on the acCD approach\cite{Aquilante_JChemPhys_Unbiased_2007} 
with a decomposition threshold of 10$^{-4}$ a.u.~as implemented in the MOLCAS 8.0 program\cite{Aquilante_JComputChem_Molcas_2015}. 
All DMRG reference wave functions, four-body RDMs, and transition three-body RDMs were calculated with the
\textsc{QCMaquis}\cite{Dolfi2014,kell15,kell16,knec16} program with its MOLCAS
interface\cite{yma16}.

Additional CCSD(T) calculations with conventional integrals were carried out with MOLCAS 8.0.
	
	\section{Results and Discussion}

	\subsection{Harvey's model 2 complex}
	
	The \emph{model 2} complex is one of the three heme models employed by \citet{Strickland_JPhysChemB_SpinForbidden_2007} in their {\it ab initio} and DFT study of ligand binding to heme, and the largest model treated with the CCSD(T) method in their work. 
	Given its small size, which still allows employing the conventional integral implementation for comparison with the CD approach,
	an electronic structure typical for SCOs and the availability of CCSD(T) reference data, we chose the \emph{model 2} complex to study the performance and the accuracy of CD-DMRG-NEVPT2 for spin-state energetics.
	
	\paragraph{Spin State Energy Differences}
	Table~\ref{tab:model2-energies} lists electronic energies of the lowest singlet ($^1A_1$) and triplet ($^3B_1$) states of the \emph{model 2} relative to the lowest quintet state ($^5B_2$) calculated with the strongly-contracted (SC-NEVPT2) and the partially-contracted (PC-NEVPT2) approach for different values of the
number of renormalized block states $m$. CCSD(T) data are presented from both our calculations and Ref.~\citenum{Strickland_JPhysChemB_SpinForbidden_2007}.
	
	The relative energies computed with SC-NEVPT2 agree well with the CCSD(T) results obtained with the
same (ANO-RCC-VDZP) basis set: the largest deviation from the CCSD(T) results is 1.3 kcal/mol. Moreover, the
variation of the relative energies with the $m$ value is even smaller (with the largest deviation being below the
chemical-accuracy threshold of 1 kcal/mol), providing further evidence that one may save computational time by using relatively
low $m$ values (evaluation of a 4-RDM scales as $m^3$)\cite{Guo_JChemTheoryComput_NElectron_2016} without significant 
accuracy loss. The difference of the SC-NEVPT2 relative energies obtained with and without CD is less than
0.1 kcal/mol for the quintet-triplet gap and about 0.3 kcal/mol for the quintet-singlet gap and therefore
comparable to the variation of the gaps with $m$. 

	
	\begin{table}[h]
		\centering
		\caption{Electronic energies of the singlet and triplet states of the \emph{model 2} complex (in kcal/mol, relative to the quintet $^5B_2$ state)}\label{tab:model2-energies}
		\begin{tabular}{cccccc}
			\toprule
			& \multicolumn{5}{c}{DMRG-SC-NEVPT2} \\
			\cline{2-6}
			$m=$ & 256* & 256 & 512 & 512/256 & 1024/256  \\
			\colrule
			$^3B_1$ & 3.4 & 3.4 & 3.1 & 2.6  & 1.9 \\ 
			$^1A_1$ & 32.8 & 33.1& 33.8& 32.8 & 31.8\\
			\colrule
			& \multicolumn{5}{c}{DMRG-PC-NEVPT2} \\
			\cline{2-6}
			$^3B_1$ & 6.5 & -34.1 & 4.1 & -1.3 & 0.5\\
			$^1A_1$ & -6.0 &  3.3 & 31.8 & 28.7 & 31.0\\
			\colrule
			& \multicolumn{3}{c}{CCSD(T)} \\
			\cline{2-6}
			& \footnote{This work, ANO-RCC-VDZP basis set} & \footnote{Ref.~\citenum{Strickland_JPhysChemB_SpinForbidden_2007}, cc-pVTZ/cc-pVDZ basis set for Fe/other atoms} & \footnote{Ref.~\citenum{Strickland_JPhysChemB_SpinForbidden_2007}, cc-pV$\infty$Z extrapolation/cc-pVDZ basis set for Fe/other atoms} \\
			\colrule
			$^3B_1$ & 3.1 & 3.7   & -2.0 \\
			$^1A_1$ & 31.5 & 25.9 & 19.6 \\
			\botrule
		\end{tabular}
	\end{table}

	\begin{figure}[h]
		\centering
		\includegraphics{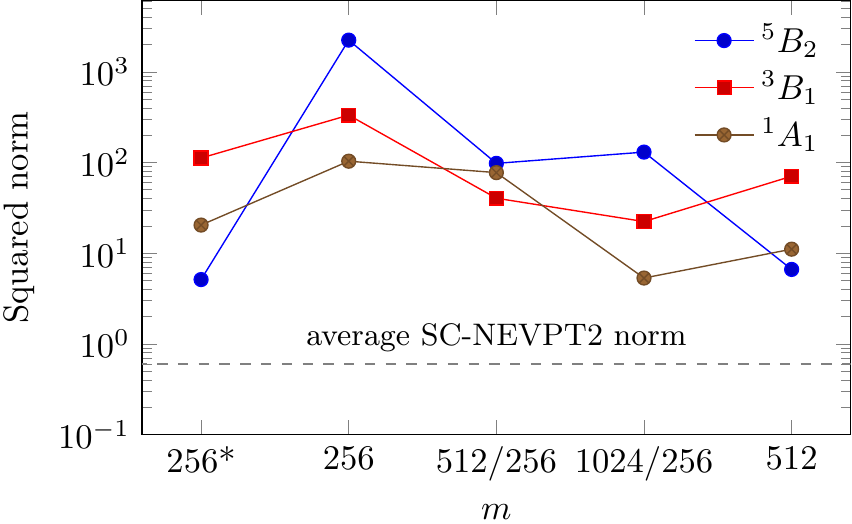}
		\caption{Squared norms of first-order wave functions of the \emph{model 2} complex calculated with DMRG-PC-NEVPT2 for different values of $m$. }\label{fig:norms}
	\end{figure}

	Unlike SC-NEVPT2, the PC-NEVPT2 relative energies are highly dependent on $m$. At $m=256$, PC-NEVPT2 predicts a
qualitatively wrong spin state ordering with the triplet state 34.1 kcal/mol below the quintet state. Only at $m=512$,
PC-NEVPT2 achieves the same spin state ordering as CCSD(T) and SC-NEVPT2. The faulty energies arise due to small
denominators in the PC-NEVPT2 energy expressions for different subspaces, which is reflected by the norm of the
PC-NEVPT2 first-order wave function shown in Fig.~\ref{fig:norms}. The norm for all states decrease by several orders of magnitude with increasing $m$, reaching from values up to 2230 ($m=256, ^5B_2$ state) down to 5 ($m=1024/256$, $^1A_1$). Although calculations without CD ($m=256$* in Fig.~\ref{fig:norms}) show lower norms than those with CD, the ordering of spin states and the relative energies are still flawed. Hence, numerical errors leading to small denominators are partially to blame on the Cholesky decomposition, but to a larger extent on the small $m$ values.
	For comparison, SC-NEVPT2 norms move in the range of $0.5$-$0.6$, regardless of the $m$ value, which is comparable with wave function norms in non-DMRG NEVPT2 calculations.\cite{Havenith_JChemPhys_Calibration_2004} No PC-NEVPT2 calculation yields a norm similar to SC-NEVPT2, although one would expect the 
PC norms to be close to the SC ones at higher $m$ values, at which, however, the four-body RDM calculations become prohibitively expensive. DMRG-PC-NEVPT2 shows therefore the problem of 'false intruder states' similar to those observed in NEVPT2 with cumulant approximations to RDMs described by \citet{Zgid_JChemPhys_study_2009}.

Concerning the dependence of PC and SC on the computational parameters,
we recall that the speed-up of the approach introduced in
the Ref.\ \citenum{Angeli_ChemPhysLett_Nelectron_2001} and fully formalized in Ref.\ \citenum{Angeli_JChemPhys_nelectron_2002}
is a consequence of the choice of the CI solution within the CAS space as the zeroth-order wave function.
This allows to greatly simplify the
formalism and to use the four-body RDMs only (otherwise also the five-body RDMs
would be required). Obviously, an approximate full CI wave function in the CAS
space (as produced by DMRG) with a formulation assuming an exact full CI wave function can 
produce numerical problems. It is furthermore 
important to stress that the 'false intruder states' are related to this inconsistency, i.e., they originate from the
approximations in the zeroth-order wave function and are unrelated to
the 'standard' intruder states.

That the PC approach is much more dependent on the quality of the zeroth-order wave function (for
instance, on the degree of CI convergence), whereas SC is more
robust and yields more stable results for a lower-quality zeroth-order
wave function can be easily understood by considering that each
denominator of SC is obtained by a weighted average of a set of
denominators in PC (the weights being related to the interactions of the
PC perturbers with the zeroth-order wave function). Therefore, large variations of the PC
denominators for perturbers with a small interaction with the zeroth-order wave function (which
can have important effects on the PC energy if the denominators approach
zero) have almost no effects on SC. 
	
	The discrepancy of our CCSD(T)/ANO-RCC-VDZP results with the CCSD(T) results of \citet{Strickland_JPhysChemB_SpinForbidden_2007} (Table~\ref{tab:model2-energies}) might point to a large basis set effect at a first glance. However, the discrepancy may also arise due to the convergence of the single-reference wave function to a different state. In our calculations, the guess orbitals for the reference ROHF wave function for the quintet and triplet CCSD(T) calculation were generated from state-averaged CASSCF calculations for three quintet states and triplet states to ensure the correct state character (and similarly in the NEVPT2 calculations). The CASSCF calculations show gaps of 2-6\,kcal/mol between the lowest quintet or triplet states, thus, a black-box single-reference calculation may well converge to an adjacent state, which will affect the spin-state energetics. This emphasizes the need for multireference methods even in cases where single-reference methods perform well.
	

%
%
	
%

\begin{figure*}[t]
			\centering
			\includegraphics[width=0.99\textwidth]{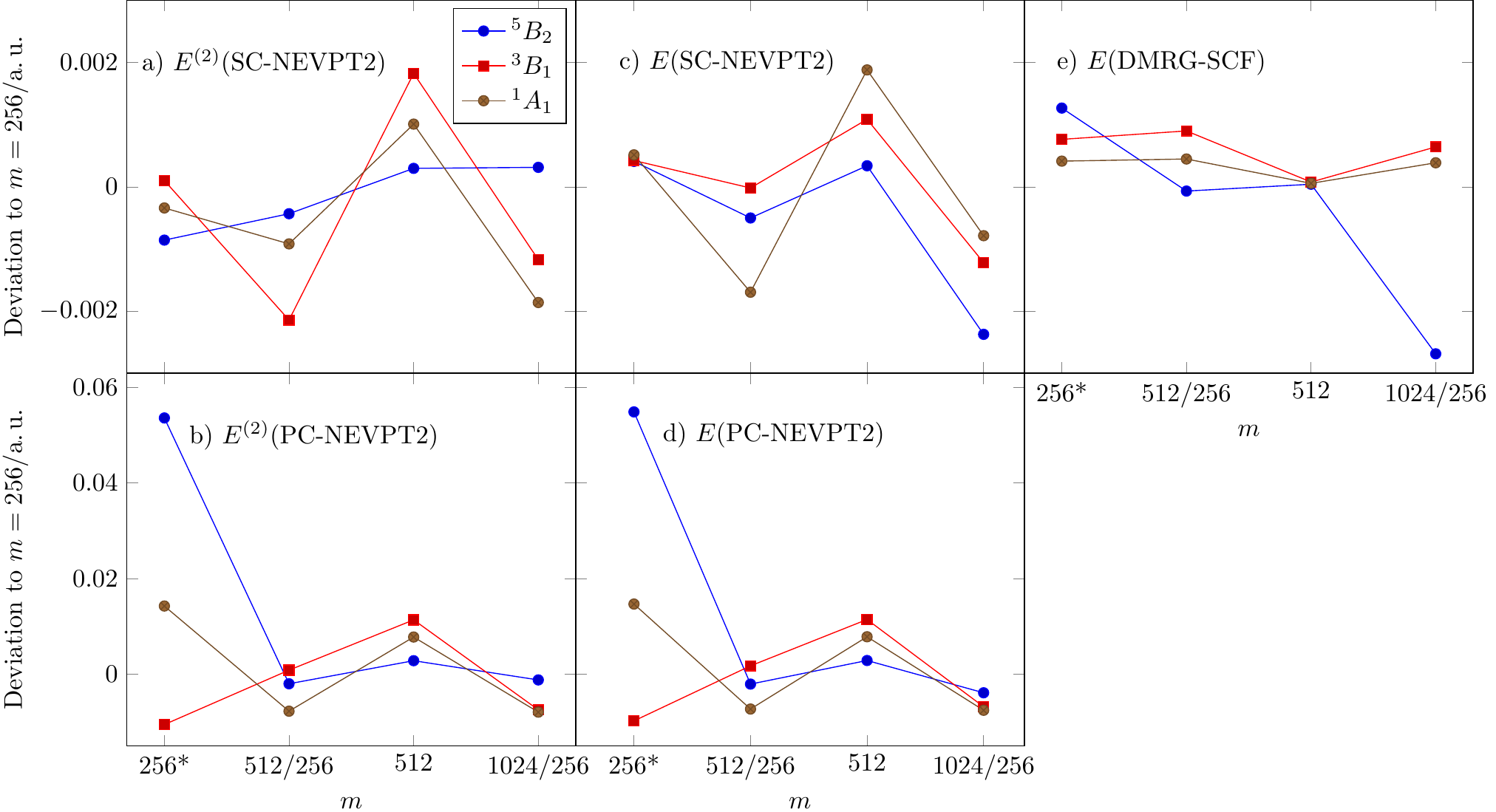}
			\caption{Differences of second-order energy corrections (a,b) and total electronic energies (c,d) obtained with DMRG-SC-NEVPT2 and DMRG-PC-NEVPT2 for various states for the \emph{model 2} complex for varying values of the number of renormalized block states $m$ compared to the results obtained with $m=256$. DMRG-SCF energy differences are shown in (e). 
}
\label{fig:abs-errors}
		\end{figure*}

	\paragraph{CD accuracy: absolute energies.} Fig.~\ref{fig:abs-errors} shows deviations of the second-order energy corrections and the total
electronic CD-DMRG-SC-PC-NEVPT2 and CD-DMRG-PC-NEVPT2 energies calculated with different $m$ values. The first point ($m=256$*) in each
subplot is simply the difference between CD and non-CD energies for $m=256$. For SC-NEVPT2, the errors in the
second-order energy arising due to Cholesky decomposition are on the order of 10$^{-4}$ atomic units and are
comparable to those (and even slightly smaller) of DMRG-SCF. The total NEVPT2 energies show even smaller errors on
average due to the cancellation of DMRG-SCF and second-order energy errors. In all cases, errors arising due to CD are
smaller than errors due to smaller $m$ values in SC-NEVPT2, which justifies CD as an approximation
for (DMRG)-SC-NEVPT2 calculations. For PC-NEVPT2, the CD errors are much larger and reach 0.05 atomic units for the quintet state, because of the numerical errors leading to 'false intruder states' mentioned earlier. However, errors associated with different $m$ values are also larger than for SC-NEVPT2, likely for the same reason. Therefore, we conclude that PC-NEVPT2 is much more prone to numerical problems arising from both Cholesky decomposition and DMRG.

	\subsection{Cobalt tropocoronand complex}
	
	Tropocoronand ligands (Fig.~\ref{fig:structures}b) with different lengths of the alkylidene chain $n$ provide a fine-tuned constrained coordination environment that varies with $n$, yielding metal complexes with interesting geometric and electronic structures, especially if coordinated with a non-innocent ligand\cite{Jorgensen_CoordChemRev_Differences_1966,Kaim_EurJInorgChem_Shrinking_2012} such as NO. \cotc{}, the smallest member of a homologous series of cobalt complexes with $n=3$ to $5$, was previously reported to be paramagnetic, unlike its higher homologs and other cobalt nitrosyl complexes\cite{Ellison_InorgChem_TiltAsymmetry_1998,Jaworska_ChemPhys_Theoretical_2007}, which contradicts a recent DFT and experimental study\cite{Hopmann_InorgChem_SingletTriplet_2015}. Metal-nitrosyl bonds are known to show a large amount of static correlation (see, e.g., Ref.\ \citenum{Freitag_PhysChemChemPhys_Orbital_2015}), which cannot be reliably described by single-reference methods such as DFT. Due to this fact and the controversy on the spin state, we study with DMRG-SCF and CD-DMRG-NEVPT2 the electronic structure and spin-state energetics of \cotc{}.
	
	\paragraph{Electronic structure of the complex.}
	In metal nitrosyl complexes it is usually not trivial to unambiguously describe the electronic structure of the metal
and the NO ligand separately by assigning them distinct electronic occupations: thus, it is common to use the
Enemark-Feltham notation\cite{Enemark_JAmChemSoc_Stereochemical_1974} to describe their electronic structure. \cotc{} is
a \{CoNO\}$^8$ complex, according to the notation, where 8 is the total number of electrons in the metal $d$ and NO
$\pi^{\ast}$ orbitals. 
	
\begin{figure*}[t]
				\centering
				\includegraphics[width=0.95\textwidth]{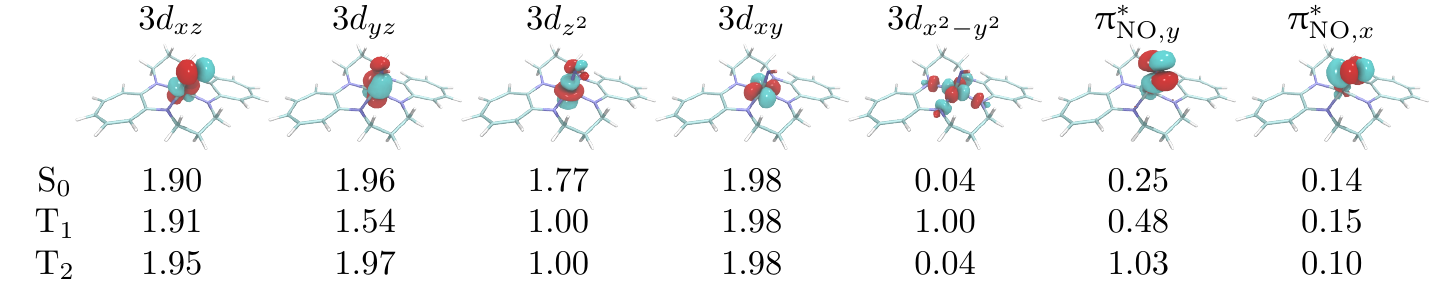}
				\caption{Natural orbital occupation numbers for orbitals participating in the Co-nitrosyl bond in \cotc{} in different spin states.}\label{fig:no-occ-numbers}
			\end{figure*}

	To gain insights into the electronic structure of the complex, we analyzed the DMRG-SCF wave functions of the
spin states, in particular the natural orbital occupation numbers of metal $d$ and NO $\pi^{\ast}$ orbitals, which are
essential for the Co--NO bond description. All results are compiled in Fig.~\ref{fig:no-occ-numbers}. 
	
	Although the orbitals are delocalised and several occupation numbers severely deviate from 2 or 0 (which is a sign of
a multireference character and significant static correlation typical for metal nitrosyl
complexes)\cite{Freitag_PhysChemChemPhys_Orbital_2015}, we may approximately assign electronic
occupations to both metal and NO. The occupation numbers in the lowest singlet state S$_0$ most closely resemble the following occupation pattern:
	
	\[
		(3d_{xz})^2 (3d_{yz})^2 (3d_{z^2})^2 (3d_{xy})^2 (3d_{x^2-y^2})^0 (\pi^{\ast}_{\text{NO},y})^0 (\pi^{\ast}_{\text{NO},x})^0
	\]
	This may be interpreted as a $d^8$ Co coupled to a NO$^+$ cation, although the occupation numbers of the $3d_{z^2}$
and $\pi^{\ast}_{\text{NO},y}$ shows a significant admixture of a neutral NO and $d^7$ Co. Analogously, the T$_2$ state may
be characterized as $d^7$ Co and neutral NO. However, the T$_1$ state shows occupation numbers of $3d_{yz}$ and
$\pi^{\ast}_{\text{NO},y}$ orbitals close to 1.5 and 0.5, which represents a situation exactly between Co $d^7$ and $d^8$
and neutral and cationic NO. The state characters are in contrast to those found in the DFT study by
\citet{Hopmann_InorgChem_SingletTriplet_2015}, where the T$_1$ and T$_2$ states are characterized as $d^7$ Co and neutral NO and as $d^6$ Co and anionic NO$^-$, respectively. However such discrepancies are not surprising as the single-configuration nature of standard Kohn-Sham DFT does not allow for an accurate description of states with multireference character. It is well known that spin densities in metal nitrosyl complexes obtained with DFT and multiconfigurational methods may differ significantly.\cite{Radon_JPhysChemB_Electronic_2010,bogu11,Boguslawski_JChemTheoryComput_Accurate_2012}
	
	\paragraph{Singlet-triplet gap.} Having characterized the electronic structure of \cotc{} with DMRG-SCF, we
calculated the S$_0$-T$_1$ energy gap with CD-DMRG-SC-NEVPT2, listed in Table~\ref{tab:cotc}. CD-DMRG-SC-NEVPT2
predicts S$_0$\ as the ground state of \cotc{}, just as DMRG-SCF and all DFT functionals do, which is consistent with the diamagnetism of its ground state recently confirmed in Ref.~\citenum{Hopmann_InorgChem_SingletTriplet_2015}. However, the gap predicted by NEVPT2 is by at least 10 kcal/mol larger than those calculated with DFT. Interestingly, the DMRG-SCF result deviates from the NEVPT2 result by only 3.6 kcal/mol, indicating that the (22,22) active space is able to capture both static and a significant portion of dynamic correlation equally well in both states.
	
	\begin{table}[h]
		\centering
		\caption{S$_0$-T$_1$ energy gap (in kcal/mol) of \cotc{} calculated with CD-DMRG-SC-NEVPT2 (abbreviated as 'SC-NEVPT2') and other methods. DFT results are taken from Ref.~\citenum{Hopmann_InorgChem_SingletTriplet_2015}.}\label{tab:cotc}
		\begin{tabular}{ccccc}
			\toprule
			SC-NEVPT2 & DMRG-SCF & OLYP\cite{Hopmann_InorgChem_SingletTriplet_2015} & PW91\cite{Hopmann_InorgChem_SingletTriplet_2015} & B3LYP-D3\cite{Hopmann_InorgChem_SingletTriplet_2015} \\
			35.0	& 38.6	& 23.8	& 25.1	& 10.4 \\
			\botrule
		\end{tabular}
	\end{table}

	\section{Conclusions}
	In this work, we have presented an implementation of the second-order \textit{N}-electron valence state perturbation theory
(NEVPT2) employing a density-matrix renormalization group (DMRG) reference wave function and Cholesky decomposition (CD)
for the two-electron repulsion integrals dubbed as CD-DMRG-NEVPT2 which is a parameter-free multireference perturbation theory applicable to large
systems. 

A multi-reference pertubation theory faces two challenges when applied to large systems. These are the calculation of the
higher-order reduced density matrix and the calculation of a large number of integrals. In this work, we have 
considered a solution to the second one, yielding a method that is capable of dealing with systems containing more than 1000
basis functions. We demonstrated the application of the method on two examples of SCOs: a smaller heme model (\emph{model 2}) and a larger cobalt nitrosyl tropocoronand complex. The strongly-contracted (SC) variant of CD-DMRG-NEVPT2 describes the spin-state energetics with a similar accuracy as CCSD(T) in the heme model, and is very insensitive to the number of renormalized block states $m$. The approximations introduced by CD turn out to be negligible. Unlike the SC variant, the partially-contracted CD-DMRG-NEVPT2 is prone to 'false intruder states'\cite{Zgid_JChemPhys_study_2009} due to numerical approximations introduced by the DMRG and Cholesky decomposition. 
	
	Subsequently, we employed a large-scale DMRG-SCF and a strongly-contracted CD-DMRG-NEVPT2 calculation to describe the electronic structure of a cobalt nitrosyl tropocoronand complex. The electronic structure of the lowest singlet and triplet states calculated with DMRG-SCF showed significant static correlation. The NEVPT2 calculation confirmed the singlet ground state and with it the diamagnetism of the complex shown experimentally in a recent study. 
Combined with the recent advances to overcome the bottleneck of the higher-order RDM evaluation (loc. cit.), we believe that
CD-DMRG-NEVPT2 will be a valuable tool in transition metal and bioinorganic chemistry for calculating energies and properties of 
large molecular systems that are governed by static electron correlation.
	
	\section*{Acknowledgments}
LF acknowledges the Austrian Science Fund FWF for a Schr\"{o}dinger fellowship (project no.\ J 3935).
This work was supported by the Swiss National Science Foundation SNF.


	\section*{References}
%

\end{document}